\documentclass[aps,prd,superscriptaddress,showpacs,floatfix, nofootinbib,preprintnumbers,twocolumn]{revtex4-1}
\usepackage[colorlinks=true, pdfstartview=FitV, linkcolor=black,citecolor=blue, urlcolor=blue,
bookmarks=false]{hyperref}
\usepackage{array}
\usepackage{epsfig}
\usepackage{amsmath}
\usepackage{multirow}
\usepackage{graphicx}
\usepackage{amsmath}
\usepackage{feynmf}
\usepackage{comment}
\usepackage[normalem]{ulem}  
\usepackage[dvips]{color} 

\renewcommand\sout{\bgroup \color{red} \ULdepth=-.5ex \ULset}

\newcommand{\vev}[1]{\langle{#1}\rangle}

\newcommand{\Log}{\log\frac{\mu^2}{m^2}}
\newcommand{\Logq}{\log\frac{Q^2}{m^2}}
\newcommand{\Logf}{\log\frac{Q^2}{\mu^2}}

\begin{document}
\title{Light vector correlator in medium: Wilson coefficients up to dimension 6 operators }

\author{HyungJoo Kim}
\affiliation{Department of Physics and Institute of Physics and Applied Physics, Yonsei University, Seoul 120-749, Korea}
\email{hugokm0322@gmail.com}
\author{Philipp Gubler}
\affiliation{Department of Physics, Keio University, Kanagawa 223-8522, Japan}
\affiliation{Research and Education Center for Natural Science, Keio University, Kanagawa 223-8521, Japan}
\author{Su Houng Lee}
\affiliation{Department of Physics and Institute of Physics and Applied Physics, Yonsei University, Seoul 120-749, Korea}
\date{\today}
\begin{abstract}
As an improvement of the QCD sum rule method to study modifications of light vector mesons in nuclear matter and/or at finite temperature, 
we calculate the Wilson coefficients of all independent gluonic non-scalar operators up to dimension 6 in 
the operator product expansion (OPE) of the vector channel for light quarks. 
To obtain the gluon part of the light quark OPE from the heavy quark one, 
we also compute the heavy quark expansion of the relevant quark condensates. 
Together with the results for the quark operators that are already available in the literature, 
this completes the OPE of the vector channel in a hot or dense medium for operators up to dimension 6. 
\end{abstract}


\maketitle

\section{Introduction}
Vector mesons in hot and/or dense matter have been in the focus of theoretical and experimental interest 
already for many years, primarily because some of them (quarkonia) are expected to be indicators of the quark-gluon plasma in heavy-ion 
collisions \cite{Matsui:1986dk} while others (light vector mesons, such as $\rho$, $\omega$ and $\phi$) were predicted to be probes of the 
partial restoration of chiral symmetry in nuclear matter \cite{Hatsuda:1991ez}. Furthermore, due to their decay into di-leptons, which do 
not interact strongly, they provide a relatively clear experimental signal that does not get distorted by the strongly interacting surrounding 
matter and therefore in principle allow direct access to the medium to be studied. 

The results to be presented in this paper will be relevant for the light vector mesons. 
Their behavior in nuclear matter has been studied intensively in various experiments during the last two decades and a 
considerable amount of new information has been obtained \cite{Hayano:2008vn,Leupold:2009kz,Metag:2017ixh}. About the $\rho$ meson, the reached consensus 
seems to point to the conclusion that its peak, which is already rather broad in vacuum, is broadened further due to nuclear matter effects 
and receives only a small mass shift \cite{Nasseripour:2007aa,Wood:2008ee}, which is in any case difficult to constrain because of the broadness 
of the peak (see, however, Ref.\,\cite{Naruki:2005kd}). 
About the $\omega$ meson, new results have emerged during the last few years, experimentally determining both the mass shift and width 
of the $\omega$ peak at normal nuclear matter density with relatively high precision \cite{Metag:2011ji,Thiel:2013cea,Friedrich:2014lba,Metag:2015lza,Kotulla:2008aa}. 
Recently, 
even the momentum dependence of the $\omega$ width in nuclear matter has been measured \cite{Friedrich:2016cms}. Finally, even though somewhat scarce, some 
results are also available for the $\phi$ meson, for which both width and mass shift at normal nuclear matter density have 
been  obtained \cite{Muto:2005za,Sakuma:2006xc}. 
Moreover, the $\phi$ nuclear transparency ratio including its momentum dependence has also been measured \cite{Polyanskiy:2010tj,Hartmann:2012ia}.
In the E16 experiment to be performed at the J-PARC facility in Tokai, the width and mass modification of the $\phi$ 
are planned to be measured again with improved precision. Moreover, the momentum dependences of these quantities will 
also be studied \cite{Aoki:2015qla}. 
For accurately interpreting these past and future experimental results, it  
is highly desirable to have a thorough theoretical understanding of how vector mesons are modified in nuclear matter.  

As one possible approach, 
the QCD sum rule method \cite{Shifman:1978bx} allows one to study the modifications of the vector mesons at both finite temperature and 
density directly from the first principles of QCD \cite{Furnstahl:1989ji,Hatsuda:1992bv,Drukarev:1988kd,Hatsuda:1991ez}. 
For recent works along this line, see Refs.\,\cite{Gubler:2011ua,Hilger:2010cn,Suzuki:2012ze,Gubler:2014pta,Hohler:2013eba,Gubler:2015yna,Kim:2015xna,Gubler:2016itj,Gubler:2016hnf,Kim:2017pos,Araki:2017ebb}.
In all QCD sum rule studies of vector mesons, the results of the 
operator product expansion (OPE) of the vector channel represent the starting point of any analysis. 
To compute the OPE in hot and/or dense matter, one needs to take into account non-scalar operators because the 
existence of a medium breaks Lorentz symmetry. 
Somewhat surprisingly,  the contributions of gluonic non-scalar operators to the OPE of the vector 
channel for light quarks up to dimension 6 have never been obtained, even though some attempts were made in Ref. \cite{Leupold:1998bt} and 
the heavy-quark case was studied in Ref. \cite{Kim:2000kj}. 
As it is well known, one cannot simply take the $m \to 0$ limit of the heavy quark OPE when switching to the light 
quark case, because some of the gluonic contributions in the heavy quark OPE become part of the quark condensate in the 
light quark limit \cite{Shifman:1978bx,Generalis:1983hb,Zschocke:2011aa,Bagan:1985zp} and therefore need to be subtracted. 
Such gluonic contributions can be obtained by performing a sort of heavy-quark expansion on the quark condensates, 
as it was discussed long ago in Ref. \cite{Generalis:1983hb} for the scalar part of the OPE and more recently in 
Ref. \cite{Zschocke:2011aa} in relation to the heavy-light-quark pseudoscalar channel (see also \cite{Buchheim:2014uda,Buchheim:2015yyc}). 
We will follow the same approach and therefore first start from the OPE of non-scalar quark operators, which can (mostly)  
be found in the literature \cite{Lee:1993ww,Leupold:1998bt}, compute the gluonic contributions of these operators via the 
heavy-quark expansion and finally subtract these from the heavy quark results of Ref. \cite{Kim:2000kj}. 
By doing this, we compute all gluonic non-scalar contributions to the vector channel OPE 
up to dimension 6. 

This paper is organized as follows. 
In section II, we provide the basic definitions of the entities to be discussed in our work. 
In section III, the OPE for quark operators will be recapitulated, after which the heavy quark expansion for 
these operators will be discussed. The section concludes with the final 
OPE result for the gluonic condensates up to dimension 6.  
Section IV is devoted to the summary and conclusions. 
For the interested reader, we briefly explain in Appendix A the spin decomposition of gluonic operators 
in $D$ dimensions that we have used in this study. 

\section{Definitions}
We start from the following correlation function for the vector current $j_{\mu}(x)=\bar{q}(x)\gamma_{\mu}q(x)$, expand the OPE results as a function of small quark mass, and keep operator terms up to mass dimension 6. 
\begin{align}
\Pi_{\mu\nu}(q)&=i\int d^{4}x e^{iqx}\langle T\{j_{\mu}(x)j_{\nu}(0)\} \rangle \\
&=\Pi_{\mu\nu}^{scalar}+\Pi_{\mu\nu}^{4,2}+\Pi_{\mu\nu}^{6,2}+\Pi_{\mu\nu}^{6,4}
\label{eq:eq2}
\end{align}
In general, we can classify the correlation function according to spin and dimension of operators which occur in the OPE. $\Pi_{\mu\nu}^{scalar}$ denotes the contributions from the scalar operators. Among the superscripts of $\Pi_{\mu\nu}^{i,j}$, the first index denotes the dimension and the second one the spin of the corresponding operators. 
From Lorentz covariance and due to the fact that the vector current is conserved, 
it can easily be shown that each term in Eq.\,(\ref{eq:eq2}) satisfies the following Lorentz structure \cite{Kim:2000kj} within the large $Q^2=-q^2$ region. 
\begin{small}
\begin{align}
\Pi_{\mu\nu}^{scalar}&=(q_{\mu}q_{\nu}-q^2 g_{\mu\nu})\Pi^{scalar}, \\
\Pi^{4,2}_{\mu\nu}{\quad\,}&=\frac{1}{Q^2}[I^{4,2}_{\mu\nu}+\frac{1}{Q^2}(q_{\rho}q_{\mu}I^{4,2}_{\rho\nu}+q_{\rho}q_{\nu}I^{4,2}_{\rho\mu})\nonumber\\
&+g_{\mu\nu}\frac{q_{\rho}q_{\sigma}}{Q^2}J^{4,2}_{\rho\sigma}+\frac{q_{\mu}q_{\nu}q_{\rho}q_{\sigma}}{Q^4}(I^{4,2}_{\rho\sigma}+J^{4,2}_{\rho\sigma})],\\
\Pi^{6,2}_{\mu\nu}{\quad\,}&=\frac{1}{Q^4}[I^{6,2}_{\mu\nu}+\frac{1}{Q^2}(q_{\rho}q_{\mu}I^{6,2}_{\rho\nu}+q_{\rho}q_{\nu}I^{6,2}_{\rho\mu})\nonumber\\
&+g_{\mu\nu}\frac{q_{\rho}q_{\sigma}}{Q^2}J^{6,2}_{\rho\sigma}+\frac{q_{\mu}q_{\nu}q_{\rho}q_{\sigma}}{Q^4}(I^{6,2}_{\rho\sigma}+J^{6,2}_{\rho\sigma})],\\
\Pi^{6,4}_{\mu\nu}{\quad\,}&=\frac{q_{\kappa}q_{\lambda}}{Q^6}[I^{6,4}_{\kappa\lambda\mu\nu}+\frac{1}{Q^2}(q_{\rho}q_{\mu}I^{6,4}_{\kappa\lambda\rho\nu}+q_{\rho}q_{\nu}I^{6,4}_{\kappa\lambda\rho\mu})\nonumber\\
&+g_{\mu\nu}\frac{q_{\rho}q_{\sigma}}{Q^2}J^{6,4}_{\kappa\lambda\rho\sigma}+\frac{q_{\mu}q_{\nu}q_{\rho}q_{\sigma}}{Q^4}(I^{6,4}_{\kappa\lambda\rho\sigma}+J^{6,4}_{\kappa\lambda\rho\sigma})].
\label{OPEstructure}
\end{align}
\end{small}
Here and throughout the whole paper, we use the following convention for the summation of Lorentz indices: $A_{\mu\mu}=\sum{g_{\mu\nu}A^{\mu\nu}}$.
While the heavy quark OPE of the correlator, $\Pi^{h.q.}_{\mu\nu}$, has only gluon condensate contributions, the light quark OPE generally has both  gluon and quark condensates, $\Pi^{l.q.}_{\mu\nu}=\Pi_{\mu\nu}^\textbf{Q}+\Pi_{\mu\nu}^\textbf{G}$. The bold superscripts $\textbf{Q}$ and $\textbf{G}$ here represent quark and gluon condensate parts of the light quark OPE, respectively. 

For the condensates appearing in this work, we use the following notations.
\begin{small}
\begin{align}
&\textbf{Quark condensates}\nonumber\\
\vev{\bar{q}jq} \equiv& \vev{g\bar{q}\gamma_{\mu}(D_{\nu}G_{\mu\nu})q}, \\
\vev{j_5^2} \equiv& \vev{g^2\bar{q} t^a \gamma_5 \gamma_{\mu}q \bar{q} t^a \gamma_5 \gamma_{\mu}q}, \\
A_{\alpha\beta}\equiv&\vev{g\bar{q}(D_{\mu}G_{\alpha\mu})\gamma_{\beta}q|_{ST}}, \\
B_{\alpha\beta}\equiv&\vev{g\bar{q}\{iD_{\alpha},\tilde{G}_{\beta\mu}\}\gamma_{5}\gamma_{\mu}q|_{ST}}, \\
C_{\alpha\beta}\equiv&\vev{m\bar{q}D_{\alpha}D_{\beta}q|_{ST}}, \\
F_{\alpha\beta}\equiv&\vev{\bar{q}\gamma_{\alpha}iD_{\beta}q|_{ST}}, \\
H_{\alpha\beta}\equiv& \vev{g^2\bar{q} t^a \gamma_5 \gamma_{\alpha}q \bar{q} t^a \gamma_5 \gamma_{\beta}q|_{ST}}, \\
K_{\alpha\beta\gamma\delta}\equiv&\vev{\bar{q}\gamma_{\alpha}D_{\beta}D_{\gamma}D_{\delta}q|_{ST}} \\
\nonumber
\end{align}
\begin{align}
&\textbf{Gluon condensates}\nonumber\\
\vev{G^2} \equiv& \vev{g^2 G^a_{\mu\nu}G^a_{\mu\nu}},\\
\vev{G^3} \equiv& \vev{g^3f^{abc} G^a_{\mu\nu}G^b_{\nu\lambda}G^c_{\lambda\mu}}, \\
\vev{j^2}\equiv&\vev{g^2(D_{\mu}G^a_{\alpha\mu})(D_{\nu}G^a_{\alpha\nu})},\\
G_{2\alpha\beta} \equiv& \vev{g^2 G^a_{\alpha\mu}G^a_{\beta\mu}|_{ST}},\\
X_{\alpha\beta}\equiv&\vev{g^2 G^a_{\mu\nu}D_{\beta}D_{\alpha}G^a_{\mu\nu}|_{ST}}, \label{eq:op1} \\
Y_{\alpha\beta}\equiv&\vev{g^2 G^a_{\alpha\mu}D_{\mu}D_{\nu}G^a_{\beta\nu}|_{ST}}, \label{eq:op2} \\
Z_{\alpha\beta}\equiv&\vev{g^2 G^a_{\alpha\mu}D_{\beta}D_{\nu}G^a_{\mu\nu}|_{ST}}, \label{eq:op3} \\
G_{4\alpha\beta\gamma\delta}\equiv&\vev{g^2 G^a_{\alpha\mu}D_{\delta}D_{\gamma}G^a_{\beta\mu}|_{ST}}
\label{OperatorDefinition}
\end{align}
\end{small}
Here, $m$ is quark mass corresponding to the $q(x)$ field. $\tilde{G}_{\alpha\beta}=\frac{1}{2}\epsilon_{\alpha\beta\mu\nu}G_{\mu\nu}$ and conventions for $\epsilon_{0123}$ and $\gamma_5$ follow those of Peskin \cite{Peskin:1995ev}. Note that therefore the operator $B_{\alpha \beta}$ is defined with an opposite sign compared to Ref.\,\cite{Lee:1993ww}, where a different 
convention was used. 
$t^a = \lambda^a/2$, where $\lambda^a$ are the Gell-Mann matrices. 
The subscript $ST$ means that the Lorentz indices are made symmetric and traceless. That is, non-scalar operators appearing in this work are categorized according to their twist determined as $twist=dimension - spin$.  Let us mention here that generally one can 
construct more gluonic condensates with twist-4, such as $\vev{g^3 f^{abc} G^{a}_{\mu \kappa} G^{b}_{\nu \lambda} G^{c}_{\kappa \lambda}|_{ST}}$. 
It was however shown in Refs.\,\cite{Kim:2000kj,Kim:2015ywa} that by the 
use of the equations of motion all of them can be reduced to the three operators [Eqs.\,(\ref{eq:op1}-\ref{eq:op3})] shown here, 
whose anomalous dimensions were only recently calculated in Ref.\,\cite{Kim:2015ywa}.
The above list is therefore complete. 

\section{OPE calculation and results}
\subsection{Light quark OPE for quark condensates}
In this work, we calculate the OPE for quark operators up to dimension 6. After taking the small quark mass limit, which means that we expand the quark propagators for small $m^2/q^2$ and keep all terms up to total mass dimension 6, we get,  
\begin{small}
\begin{align}
\bigg\lbrace \Pi^\textbf{Q}_{\mu\nu} \bigg\rbrace : \nonumber\\
\Pi^{scalar}&=\frac{2m\vev{\bar{q}q}}{Q^4}-\frac{8m^3 \vev{	\bar{q}q}}{3Q^6}-\frac{4\vev{\bar{q}jq}}{9Q^6}
-\frac{2\vev{j_5^2}}{Q^6}, \label{eq:quark.ope.1}\\
I^{4,2}_{\mu\nu}&=(4-15\frac{m^2}{Q^2})F_{\mu\nu},\\
J^{4,2}_{\mu\nu}&=(-4+9\frac{m^2}{Q^2}) F_{\mu\nu},\\
I^{6,2}_{\mu\nu}&=\frac{5}{2}A_{\mu\nu}-\frac{1}{2}B_{\mu\nu}-13C_{\mu\nu}+4H_{\mu\nu},\\
J^{6,2}_{\mu\nu}&=-\frac{3}{2}A_{\mu\nu}+\frac{7}{2}B_{\mu\nu}-5C_{\mu\nu}-4H_{\mu\nu},\\
I^{6,4}_{\mu\nu\kappa\lambda}&=-16 i K_{\mu\nu\kappa\lambda},\\
J^{6,4}_{\mu\nu\kappa\lambda}&=16 i K_{\mu\nu\kappa\lambda}. \label{eq:quark.ope.7}
\end{align}
\end{small}

Note that the four-quark condensate terms $\vev{j_5^2}$ and $H_{\mu\nu}$ do not 
play any role for the computations of this paper as their redefiniton in terms of the heavy quark 
expansion leads only to gluonic terms of oder $\alpha_s^2$. We therefore will not consider them 
in what follows. 
We have confirmed that the scalar result is consistent with that of 
Refs.\,\cite{Shifman:1978bx,Generalis:1983hb} and the spin-2 and spin-4 parts agree with those 
of Refs.\,\cite{Lee:1993ww,Hatsuda:1992bv,Gubler:2015uza}. 

\subsection{Heavy Quark Expansion}
Next, we need to evaluate the gluonic components of the heavy quark condensates to 
obtain the light quark OPE for gluon condensates from the heavy quark one \cite{Generalis:1983hb}. 
For this purpose, we follow the techniques of the HQE (heavy quark expansion) introduced in Ref.\,\cite{Grozin:1986xh} 
(in Ref.\,\cite{Zschocke:2011aa}, the same procedure was referred to as ``operator mixing", and was discussed as the 
relation between normal-ordered and non-normal-ordered operators).  
The essence of the method can be summarized as follows. 
In momentum space, quark bilinear condensates can be represented by a closed one loop and are calculated in Fock-Schwinger gauge as, 
\begin{small}
\begin{align}
\vev{\bar{q}O[D_{\mu}]q}=-i\int \frac{d^D p}{(2\pi)^D}\vev{\text{Tr}_{C,D}[O[-ip_{\mu}-i \tilde{A}_{\mu}]S(p)]}. 
\label{eq:hqe}
\end{align}
\end{small}
For more details such as the detailed meaning of $\tilde{A}_{\mu}$, see for instance Ref.\,\cite{Zschocke:2011aa}.
Here, the loop integral is computed in $D = 4 - \epsilon$ dimensions with dimensional regularization. 
In doing this, some care is however needed, because inconsistencies can occur 
depending on which dimension ($D$ or 4) is used for the spin decomposition of operators. 
For example, HQE's scalar part of $O_{\alpha\beta}$ and that of its scalar decomposed one in 4 dimensions, $\frac{1}{4}g_{\alpha\beta}O_{\mu\mu}$, 
will generally have a different result for non-logarithmic terms. 
The reason why this inconsistency occurs is that direct HQE using Eq.\,(\ref{eq:hqe}) involves the spin decomposition process in D dimension and not in 4. 
We have explicitly checked that all direct HQEs of general quark operators are consistent with those of their scalar parts as long 
as all calculations (including the spin decomposition) are performed in $D$ dimensions, which is the strategy that we will adopt in this work. 
For the above consistency to hold, it should be noted that the spin decomposition of gluonic operators should also be carried out in 
$D$ dimensions. Details of this procedure are given in the Appendix A. 
This treatment differs from that of Ref.\,\cite{Zschocke:2011aa}, where furthermore $\text{Tr}[I]=D$ was used in some instances to avoid the above-mentioned problem. 
We however find that this approach is not valid in every case. 
 
Employing the strategy explained in the preceding paragraph, the HQE results for the condensates appearing in this work are obtained as follows. 
\begin{scriptsize}
\begin{align}
\vev{\bar{q}q}=&-\frac{\vev{G^2}}{48\pi^2m}-\frac{\vev{ G^3}}{1440\pi^2m^3}-\frac{\vev{j^2}}{120\pi^2m^3} \label{eq:hqe.op1} \\
\vev{\bar{q}jq}=&-\frac{\vev{j^2}}{24\pi^2}\Log \label{eq:hqe.op2} \\
A_{\alpha\beta}=&\frac{Y_{\alpha\beta}}{24\pi^2}\Log \\
B_{\alpha\beta}=&\frac{m^2 G_{2\alpha\beta}}{8\pi^2}\bigg(\Log-1\bigg)-\frac{X_{\alpha\beta}}{48\pi^2}\bigg(\Log-2 \bigg)\nonumber\\
&+\frac{Y_{\alpha\beta}}{48\pi^2}\bigg(\Log-2\bigg)-\frac{Z_{\alpha\beta}}{16\pi^2}\bigg(\Log-2 \bigg) \\
C_{\alpha\beta}=&-\frac{m^2 G_{2\alpha\beta}}{48\pi^2}\Log -\frac{X_{\alpha\beta}}{240\pi^2}+\frac{Y_{\alpha\beta}}{480\pi^2}-\frac{Z_{\alpha\beta}}{480\pi^2}\\
F_{\alpha\beta}=&-\frac{ G_{2\alpha\beta}}{24\pi^2}\Log -\frac{X_{\alpha\beta}}{960\pi^2 m^2}+\frac{Y_{\alpha\beta}}{120\pi^2 m^2}-\frac{3Z_{\alpha\beta}}{160\pi^2 m^2}\\
K_{\alpha\beta\gamma\delta}=&i\frac{11}{480\pi^2}\Log G_{4\alpha\beta\gamma\delta} \label{eq:hqe.op7}
\end{align}
\end{scriptsize}
Here, we have used dimensional regularization in combination with the $\overline {\text {MS}}$ scheme. $\mu$ is the renormalization 
scale. In the above expansion, we ignored all gluonic operators with dimension larger than 6.
Note that there are a number of terms on the right hand side of these equations that diverge in the small $m$ limit. 
These terms will get canceled by respective terms of the heavy quark OPE to be discussed in the next subsection. 
Let us furthermore mention that the results for the scalar condensates [Eqs.\,(\ref{eq:hqe.op1}) and (\ref{eq:hqe.op2})] completely agree 
with those given in Ref.\,\cite{Generalis:1983hb}. 

\subsection{Light Quark OPE for Gluon condensates}
After these preparations, we can now 
obtain the light quark OPE for gluon condensate from the following formula.   
\begin{small}
\begin{align}
\Pi^{\textbf{G}}_{\mu\nu}=\lim_{\substack{{m^2}/{q^2} \rightarrow 0}}\bigg\lbrace\Pi^{h.q.}_{\mu\nu}\bigg\rbrace-\Pi^{\textbf{Q}}_{\mu\nu}|_\text{G}
\label{eq:final.formula}
\end{align}
\end{small}
The subscript $\text{G}$ in $\Pi^{\textbf{Q}}_{\mu\nu}|_\text{G}$ stands for the replacement of quark condensates into gluon ones via the heavy quark expansion. 
Each correlation function on the r.h.s has mass singularities as seen in the 
last subsection, but they should be canceled on the l.h.s. 
To use the above formula, we need the light quark limit of the heavy quark OPE, $\lim_{\substack{{m^2}/{q^2} \rightarrow 0}} \lbrace\Pi^{h.q.}_{\mu\nu}\rbrace$, 
which can easily be extracted from the formulas given in Ref.\,\cite{Kim:2000kj}. The results read
\begin{scriptsize}
\begin{align}
\bigg\lbrace \lim_{\substack{{m^2}/{q^2} \rightarrow 0}}&\Pi^{h.q.}_{\mu\nu} \bigg\rbrace :\nonumber\\
\Pi^{scalar}&=\frac{1}{\pi^2 Q^4}(-\frac{1}{48}+\frac{1}{12}\frac{m^2}{Q^2})\vev{G^2} \nonumber\\
&+\frac{1}{\pi^2 Q^6}(-\frac{1}{720}\frac{Q^2}{m^2}+\frac{1}{540})\vev{G^3}\nonumber\\
&+\frac{1}{\pi^2 Q^6}(-\frac{1}{60}\frac{Q^2}{m^2 }+\frac{41}{1620}+\frac{1}{54}\Logq)\vev{j^2}\\
I^{4,2}_{\mu\nu}&=(\frac{1}{8\pi^2}-\frac{11}{12\pi^2}\frac{m^2}{Q^2}+(-\frac{1}{6\pi^2}+\frac{5}{6\pi^2}\frac{m^2}{Q^2})\Logq)G_{2\mu\nu},\\
J^{4,2}_{\mu\nu}&=(-\frac{7}{24\pi^2}+\frac{1}{12\pi^2}\frac{m^2}{Q^2}+(\frac{1}{6\pi^2}+\frac{1}{6\pi^2}\frac{m^2}{Q^2})\Logq)G_{2\mu\nu},\\
I^{6,2}_{\mu\nu}&=(-\frac{1}{240\pi^2}\frac{Q^2}{m^2}+\frac{31}{960\pi^2}+\frac{1}{96\pi^2}\Logq)X_{\mu\nu}\nonumber\\
&+(\frac{1}{30\pi^2}\frac{Q^2}{m^2}-\frac{739}{2880\pi^2}+\frac{3}{32\pi^2}\Logq)Y_{\mu\nu}\nonumber\\
&+(-\frac{3}{40\pi^2}\frac{Q^2}{m^2}+\frac{293}{960\pi^2}+\frac{1}{32\pi^2}\Logq)Z_{\mu\nu},\\
J^{6,2}_{\mu\nu}&=(\frac{1}{240\pi^2}\frac{Q^2}{m^2}+\frac{103}{960\pi^2}-\frac{7}{96\pi^2}\Logq)X_{\mu\nu}\nonumber\\
&+(-\frac{1}{30\pi^2}\frac{Q^2}{m^2}+\frac{71}{960\pi^2}+\frac{1}{96\pi^2}\Logq)Y_{\mu\nu}\nonumber\\
&+(\frac{3}{40\pi^2}\frac{Q^2}{m^2}+\frac{29}{960\pi^2}-\frac{7}{32\pi^2}\Logq)Z_{\mu\nu},\\
I^{6,4}_{\mu\nu\kappa\lambda}&=(-\frac{133}{180\pi^2}+\frac{11}{30\pi^2}\Logq)G_{4\mu\nu\kappa\lambda},\\
J^{6,4}_{\mu\nu\kappa\lambda}&=(\frac{181}{180\pi^2}-\frac{11}{30\pi^2}\Logq)G_{4\mu\nu\kappa\lambda}.
\label{OPEGsmallmass}
\end{align}
\end{scriptsize}
Note that, by independently checking the computations of Ref.\,\cite{Kim:2000kj}, we found that 
both formulas given in Eq.\,(20) of that reference should be multiplied by a factor of $1/2$, which was already mentioned in Ref.\,\cite{Kim:2017pos}.
This fact is taken into account for the above results. 

Finally, substituting the results of Eqs.\,(\ref{eq:hqe.op1}-\ref{eq:hqe.op7}) into Eqs.\,(\ref{eq:quark.ope.1}-\ref{eq:quark.ope.7}), 
and thereafter using Eq.\,(\ref{eq:final.formula}), we see that indeed all 
mass singularities cancel and obtain the following final expression: 
\begin{scriptsize}
\begin{align}
\bigg\lbrace \Pi^{\textbf{G}}_{\mu\nu}& \bigg\rbrace :\nonumber\\
\Pi^{scalar}&=\frac{1}{\pi^2 Q^4}(\frac{1}{48}+\frac{1}{36}\frac{m^2}{Q^2})\vev{G^2}\nonumber\\
&+\frac{1}{\pi^2 Q^6}(\frac{1}{324}+\frac{1}{54}\Logf)\vev{j^2} \label{eq:final.res.1}\\
I^{4,2}_{\mu\nu}&=(\frac{1}{8\pi^2}-\frac{47}{48\pi^2}\frac{m^2}{Q^2}+(-\frac{1}{6\pi^2}+\frac{5}{6\pi^2}\frac{m^2}{Q^2})\Logf)G_{2\mu\nu},\\
J^{4,2}_{\mu\nu}&=(-\frac{7}{24\pi^2}+\frac{25}{48\pi^2}\frac{m^2}{Q^2}+(\frac{1}{6\pi^2}+\frac{1}{6\pi^2}\frac{m^2}{Q^2})\Logf)G_{2\mu\nu},\\
I^{6,2}_{\mu\nu}&=(-\frac{1}{60\pi^2}+\frac{1}{96\pi^2}\Logf)X_{\mu\nu}\nonumber\\
&+(-\frac{361}{2880\pi^2}+\frac{3}{32\pi^2}\Logf)Y_{\mu\nu}\nonumber\\
&+(\frac{19}{320\pi^2}+\frac{1}{32\pi^2}\Logf)Z_{\mu\nu},\\
J^{6,2}_{\mu\nu}&=(-\frac{1}{20\pi^2}-\frac{7}{96\pi^2}\Logf)X_{\mu\nu}\nonumber\\
&+(\frac{149}{960\pi^2}+\frac{1}{96\pi^2}\Logf)Y_{\mu\nu}\nonumber\\
&+(-\frac{239}{960\pi^2}-\frac{7}{32\pi^2}\Logf)Z_{\mu\nu},\\
I^{6,4}_{\mu\nu\kappa\lambda}&=(-\frac{133}{180\pi^2}+\frac{11}{30\pi^2}\Logf)G_{4\mu\nu\kappa\lambda},\\
J^{6,4}_{\mu\nu\kappa\lambda}&=(\frac{181}{180\pi^2}-\frac{11}{30\pi^2}\Logf)G_{4\mu\nu\kappa\lambda}.
\label{eq:final.res.7}
\end{align}
\end{scriptsize}
From these results, one can now easily extract the transverse and longitudinal part of the correlator and can 
derive the corresponding sum rules. 
While the values of scalar and some twist-2 non-scalar gluonic condensates in nuclear matter were discussed already a long time 
ago \cite{Hatsuda:1991ez,Jin:1992id}, it is also possible to give rough estimates for the twist-4 gluonic condensates which were the 
main target of this work \cite{Kim:2000kj}. 
With these estimates, it will be possible to study the consequences of our results on the behavior of vector mesons at finite density. 
We expect that the twist-4 gluonic condensates could have some non-negligible effect in particular on the modification of the vector 
meson masses at non-zero momenta, namely their dispersion relations \cite{Lee:1997zta}. 
While the effect of gluonic condensates will likely be rather small for the $\rho$ and $\omega$ channels, where the 
finite density modifications of the OPE are dominated by quark condensate terms \cite{Leupold:1998bt}, their relative importance 
will increase for the $\phi$ meson case, where finite density effects due to quark condensates are suppressed \cite{Gubler:2014pta}.
This could be relevant for the future interpretation of 
experimentally measured spectra, which always involve vector mesons that move with some finite velocity relative to the surrounding 
nuclear matter.    

\section{Summary and Conclusions}
In this work, we have for the first time computed the Wilson coefficients, at leading order in $\alpha_s$, of dimension 6, 
spin-2 and spin-4 gluonic operators in the OPE of the vector correlator for light quarks. 
We have also obtained the leading order $\alpha_s$ Wilson coefficient of the dimension 4, spin-2 gluon operator 
(including its $m^2$ correction) which has so far never been correctly given in the literature. 
For self-adjoint mesons, this completes the vector channel OPE for all possible scalar and non-scalar operators 
up to dimension 6 that can have non-zero expectation values in a hot and/or dense medium that is invariant under parity and 
time reversal. 

To reach our final results, given in Eqs.\,(\ref{eq:final.res.1}-\ref{eq:final.res.7}), we 
followed the (standard) procedure of starting from the OPE expression of gluonic operators for arbitrary quark masses 
(which is usually used for the OPE of the heavy quark correlator), taking its small quark mass limit and subtracting from it the 
contributions that become part of the quark condensates in this limit. 
This subtraction cancels all mass singularities for $m \to 0$, which appear at the intermediate steps of the 
computation and leads to a well-behaved final expression. 

In the future, we plan to apply our results to the QCD sum rule analyses of light vector mesons 
in nuclear matter and/or at finite temperature. As the non-scalar operators that we have studied in this work affect the momentum dependence of the 
mesons in a non-trivial way (i.e. they modify the dispersion relation observed in vacuum), it will be especially interesting 
to investigate the behavior of the vector mesons with non-zero momentum and to provide post- and predictions for 
past and future experiments that measure vector mesons in nuclei.

\section*{Acknowledgements}
This work was supported by the Korea National Research Foundation under the grant number 2016R1D1A1B03930089. The research of P.G. is supported by Mext-Supported Program for the Strategic Foundation at Private Universities, ``Topological Science" under Grant No. S1511006

\appendix

\section{Spin decomposition of gluonic operators in $D$ dimensions}
In this appendix, we provide formulas of the spin decompositions of gluonic operators in $D$ dimensions used in our 
work. 
\subsection{Decomposition of $\langle g^2 G_{\kappa \alpha}^{a} G_{\lambda \beta}^{a}  \rangle$}
From the symmetry properties of $\langle g^2 G_{\kappa \alpha}^{a} G_{\lambda \beta}^{a}  \rangle$, its Lorentz structure can be decomposed into 
\begin{small}
\begin{align}
& \langle g^2 G_{\kappa \alpha}^{a} G_{\lambda \beta}^{a} \rangle = \nonumber\\
& a c_{\kappa \lambda \alpha \beta}  + g_{\kappa \lambda} b_{\alpha \beta} - g_{\kappa \beta} b_{\alpha \lambda}
+ g_{\alpha \beta} b_{\kappa \lambda} - g_{\alpha \lambda} b_{\kappa \beta}.
\label{eq:ap1}
\end{align}
\end{small}
with
\begin{small}
\begin{equation}
c_{\kappa \lambda \alpha \beta} = g_{\kappa \lambda}g_{\alpha \beta} - g_{\kappa \beta}g_{\alpha \lambda}. 
\label{eq:ap2}
\end{equation}
\end{small}

Contracting the Lorentz indices with with one or two metric tensors, the variable $a$ and the tensor $b_{\alpha \beta}$ are determined as 
\begin{small}
\begin{equation}
\begin{split}
a =& \frac{1}{(D-1)(D-2)} \langle G^2  \rangle,  \\
b_{\alpha \beta} =&  \frac{1}{D-2} G_{2 \alpha \beta}.
\end{split}
\label{eq:ap3}
\end{equation}
\end{small}

\subsection{Decomposition of $\langle g^3 f^{abc} G_{\mu \nu}^{a} G_{\alpha \beta}^{b} G_{\rho \sigma}^{c} \rangle$}
Let us first study the scalar part of  this operator. 
Its Lorentz structure can be reduced as 
\begin{small}
\begin{align}
& \langle g^3 f^{abc} G^a_{\mu \nu} G^b_{\alpha \beta} G^c_{\rho \sigma} \rangle |_{\textrm{scalar}} = \nonumber \\
& a( g_{\mu \alpha} c_{\nu \rho \beta \sigma}  
- g_{\mu \beta} c_{\nu \rho \alpha \sigma} 
- g_{\mu \rho} c_{\nu \alpha \sigma \beta} 
+ g_{\mu \sigma} c_{\nu \alpha \rho \beta}).
\label{eq:ap4}
\end{align}
\end{small}
Contracting the Lorentz indices, $a$ is determined as 
\begin{equation}
a =  \frac{1}{D(D-1)(D-2)} \langle  G^3  \rangle. 
\label{eq:ap5}
\end{equation}

Next, we consider the spin-2 part, that we represent as one general spin-2 tensor $a_{\alpha \beta}$, which is symmetric and traceless. 
Using again the symmetries of $\langle f^{abc} G_{\mu \nu}^{a} G_{\alpha \beta}^{b} G_{\rho \sigma}^{c} \rangle$, we 
obtain
\begin{align}
& \langle g^3 f^{abc} G^a_{\mu \nu} G^b_{\alpha \beta} G^c_{\rho \sigma} \rangle |_{\textrm{spin-2}} = \nonumber \\
& a_{\mu \alpha} c_{\nu \rho \beta \sigma} 
- a_{\mu \beta} c_{\nu \rho \alpha \sigma} 
- a_{\mu \rho} c_{\nu \alpha \sigma \beta} \nonumber \\
& + a_{\mu \sigma} c_{\nu \alpha \rho \beta} 
- a_{\nu \alpha} c_{\mu \rho \beta \sigma}
+ a_{\nu \beta} c_{\mu \rho \alpha \sigma} \nonumber \\
& + a_{\nu \rho} c_{\mu \alpha \sigma \beta}
- a_{\nu \sigma} c_{\mu \alpha \rho \beta}
+ a_{\alpha \rho} c_{\mu \beta \nu \sigma} \nonumber \\ 
& - a_{\alpha \sigma} c_{\mu \beta \nu \rho}
-  a_{\beta \rho} c_{\mu \alpha \nu \sigma}
+ a_{\beta \sigma} c_{\mu \alpha \nu \rho}.
\label{eq:ap6}
\end{align}
Contracting with two metric tensors and using formula (A.3) of Ref.\,\cite{Kim:2000kj}, we get
\begin{small}
\begin{align}
a_{\mu \alpha} =& \frac{1}{(D-2)(D-3)} \langle  g^3 f^{abc} G^a_{\mu \nu} G^b_{\alpha \beta} G^c_{\nu \beta} \rangle \nonumber \\
=& \frac{1}{2(D-2)(D-3)} (X_{\mu \alpha} + 2 Z_{\mu \alpha}).  
\label{eq:ap7}
\end{align}
\end{small}

\subsection{Decomposition of $\langle g^2 G^a_{\mu_1 \nu_1} D_{\alpha} D_{\beta} G^a_{\mu_2 \nu_2} \rangle$}
We start again with the scalar part, whose Lorentz structure can be reduced as  
\begin{small}
\begin{align}
& \langle g^2 G^a_{\mu_1 \nu_1} D_{\alpha} D_{\beta} G^a_{\mu_2 \nu_2} \rangle |_{\textrm{scalar}} =  
a g_{\alpha \beta} c_{\mu_1 \mu_2 \nu_1 \nu_2} \nonumber \\
&+ b ( g_{\nu_2 \alpha} c_{\mu_1 \mu_2 \nu_1 \beta} - g_{\mu_2 \alpha} c_{\mu_1 \nu_2 \nu_1 \beta}) \nonumber \\
&+ d ( g_{\nu_2 \beta} c_{\mu_1 \mu_2 \nu_1 \alpha} - g_{\mu_2 \beta} c_{\mu_1 \nu_2 \nu_1 \alpha}). 
\label{eq:ap8}
\end{align}
\end{small}
Contracting the Lorentz indices and making use of the fact that anti-symmetrizing Eq.\,(\ref{eq:ap8}) 
within the indices attached to the covariant derivatives ($\alpha$ and $\beta$) leads to an operator with three 
gluon fields that we discussed in the previous subsection, one can derive the following result for $a$, $b$ and $d$: 
\begin{small}
\begin{align}
a = & \frac{2}{(D+2)D} \Bigl( \frac{1}{D-2} \langle  G^3  \rangle - \frac{1}{D-1} \langle j^2 \rangle \Bigr) , \\
b = & \frac{1}{(D+2)D} \Bigl( \frac{1}{D-2} \langle  G^3  \rangle - \frac{1}{D-1} \langle j^2 \rangle \Bigr) , \\
d = & -\frac{1}{(D+2)D(D-1)} \Bigl( \frac{3}{D-2} \langle  G^3  \rangle + \frac{1}{D-1} \langle j^2 \rangle \Bigr). 
\label{eq:ap9}
\end{align}
\end{small}
Next, we study the spin-2 part, which we parametrize by symmetric and traceless tensors  $a^1_{\alpha \beta}$, $a^2_{\alpha \beta}$, $\dots$ 
Using the symmetries of $\langle g^2 G^a_{\mu_1 \nu_1} D_{\alpha} D_{\beta} G^a_{\mu_2 \nu_2} \rangle$, we get 
\begin{small}
\begin{align}
& \langle g^2 G^a_{\mu_1 \nu_1} D_{\alpha} D_{\beta} G^a_{\mu_2 \nu_2} \rangle|_{\textrm{spin-2}} = \nonumber \\ 
& a^1_{\mu_1 \mu_2} g_{\nu_1 \nu_2} g_{\alpha \beta} + a^1_{\nu_1 \nu_2} g_{\mu_1 \mu_2} g_{\alpha \beta} + a^2_{\alpha \beta} c_{\mu_1 \mu_2 \nu_1 \nu_2} \nonumber \\ 
&- a^1_{\nu_1 \mu_2} g_{\mu_1 \nu_2} g_{\alpha \beta} - a^1_{\mu_1 \nu_2} g_{\nu_1 \mu_2} g_{\alpha \beta} + b^1_{\mu_1 \mu_2} g_{\nu_1 \beta} g_{\nu_2 \alpha} \nonumber \\
&+ b^2_{\nu_1 \beta} c_{\mu_1 \mu_2 \alpha \nu_2} + b^2_{\nu_2 \alpha} c_{\mu_1 \mu_2 \nu_1 \beta} - b^1_{\nu_1 \mu_2} g_{\mu_1 \beta} g_{\nu_2 \alpha} \nonumber \\
&- b^2_{\mu_1 \beta} c_{\nu_1 \mu_2 \alpha \nu_2} - b^1_{\mu_1 \nu_2} g_{\nu_1 \beta} g_{\mu_2 \alpha} - b^2_{\mu_2 \alpha} c_{\mu_1 \nu_2 \nu_1 \beta} \nonumber \\
&+ b^1_{\nu_1 \nu_2} g_{\mu_1 \beta} g_{\mu_2 \alpha} + d^1_{\mu_1 \mu_2} g_{\nu_1 \alpha} g_{\nu_2 \beta} + d^2_{\nu_1 \alpha} c_{\mu_1 \mu_2 \beta \nu_2} \nonumber \\
&+ d^2_{\nu_2 \beta} c_{\mu_1 \mu_2 \nu_1 \alpha} - d^1_{\nu_1 \mu_2} g_{\mu_1 \alpha} g_{\nu_2 \beta} - d^2_{\mu_1 \alpha} c_{\nu_1 \mu_2 \beta \nu_2} \nonumber \\ 
&- d^1_{\mu_1 \nu_2} g_{\nu_1 \alpha} g_{\mu_2 \beta} - d^2_{\mu_2 \beta} c_{\mu_1 \nu_2 \nu_1 \alpha} + d^1_{\nu_1 \nu_2} g_{\mu_1 \alpha} g_{\mu_2 \beta}.
\label{eq:ap10}
\end{align}
\end{small}
Taking all possible contractions, we can derive specific expressions for the tensors $a^1_{\alpha \beta}$, $a^2_{\alpha \beta}$, $\dots$ 
These read
\begin{scriptsize}
\begin{align}
a^1_{\alpha \beta} =& \frac{1}{(D+4)(D+1)(D-2)} \Biggl[ \frac{2D^2 + D - 7}{2(D-3)} X_{\alpha \beta} \nonumber \\
& + \frac{D^2 + 3D -2}{D} Y_{\alpha \beta} +  \frac{D^3 + 3D^2 - 6D + 6}{D(D-3)} Z_{\alpha \beta} \Biggr], \\
a^2_{\mu \nu} =& \frac{1}{(D+4)(D+1)(D-2)} \Biggl[ \frac{D^2 - D + 1}{D-3} X_{\alpha \beta} \nonumber \\
& + \frac{4}{D} Y_{\alpha \beta} +  \frac{2(2D^2 + 3D - 6)}{D(D-3)} Z_{\alpha \beta} \Biggr], \\
b^1_{\alpha \beta} =& \frac{1}{D+1} \Biggl[ \frac{1}{2(D-3)} X_{\alpha \beta} \nonumber \\
& + \frac{D - 1}{D(D-2)} Y_{\alpha \beta} +  \frac{D^2 - 3D + 3}{D(D-2)(D-3)} Z_{\alpha \beta} \Biggr], \\
b^2_{\alpha \beta} =& \frac{1}{(D+4)(D+1)(D-2)} \Biggl[ \frac{D^2 - D + 1}{2(D-3)} X_{\alpha \beta} \nonumber \\
& + \frac{2}{D} Y_{\alpha \beta} +  \frac{2D^2 + 3D - 6}{D(D-3)} Z_{\alpha \beta} \Biggr], \\
d^1_{\alpha \beta} =& \frac{1}{(D+1)(D-2)} \Biggl[ -\frac{3}{2(D-3)} X_{\alpha \beta} \nonumber \\
& + \frac{D-1}{D} Y_{\alpha \beta} - \frac{4D - 3}{D(D-3)} Z_{\alpha \beta} \Biggr], \\
d^2_{\alpha \beta} =& \frac{1}{(D+4)(D+1)(D-2)} \Biggl[ -\frac{3(2D+1)}{2(D-3)} X_{\alpha \beta} \nonumber \\
& + \frac{2}{D} Y_{\alpha \beta} - \frac{D^3 + 3D^2 + D + 6}{D(D-3)} Z_{\alpha \beta} \Biggr].
\label{eq:ap11}
\end{align}
\end{scriptsize}
Note that the $D=4$ limit of this result differs from that of Eq.\,(D.4) of Ref.\,\cite{Kim:2000kj}, which 
contains several typos. 

Finally, we consider the spin-4 part of this operator, for which we need a symmetric and traceless tensor $a_{\alpha \beta \gamma \delta}$. 
Making again use of the symmetries of $\langle g^2 G^a_{\mu_1 \nu_1} D_{\alpha} D_{\beta} G^a_{\mu_2 \nu_2} \rangle$, we obtain  
\begin{small}
\begin{align}
& \langle g^2 G^a_{\mu_1 \nu_1} D_{\alpha} D_{\beta} G^a_{\mu_2 \nu_2} \rangle|_{\textrm{spin-4}}= \nonumber \\
& g_{\mu_1 \mu_2} a_{\nu_1 \nu_2 \alpha \beta} - g_{\mu_1 \nu_2} a_{\nu_1 \mu_2 \alpha \beta} \nonumber \\
& - g_{\nu_1 \mu_2} a_{\mu_1 \nu_2 \alpha \beta} + g_{\nu_1 \nu_2} a_{\mu_1 \mu_2 \alpha \beta}.  
\label{eq:ap12}
\end{align}
\end{small}
Contracting two suitable Lorentz indices, we derive 
\begin{small}
\begin{align}
a_{\alpha \beta \gamma \delta} = \frac{1}{D-2} G_{4\alpha \beta \gamma \delta}.
\label{eq:ap13}
\end{align}
\end{small}

\end{document}